\documentclass[5p,sort&compress]{elsarticle}
\usepackage{amsmath}
\usepackage{hyperref}
\usepackage[usenames, dvipsnames]{color}
\usepackage{booktabs}
\usepackage{multirow}

\usepackage[T1]{fontenc}
\usepackage{lmodern}
\usepackage{relsize}
\usepackage{tcolorbox}
\usepackage{colortbl}

\usepackage{footmisc}
\definecolor{green(html/cssgreen)}{rgb}{0.0, 0.5, 0.0}

\usepackage{multirow}
\usepackage{array}
\usepackage{booktabs}

\usepackage{enumitem}

\usepackage{amsmath,amssymb,amsfonts}
\usepackage{graphicx}
\usepackage{textcomp}
\usepackage{xcolor}
\usepackage{comment}
\usepackage{tcolorbox}
\usepackage{algorithm}
\usepackage{algpseudocode}
\usepackage{fixltx2e}
\usepackage{boldline}
\usepackage{bm}
\usepackage{makecell}
\usepackage{multirow}
\usepackage{subcaption}
\usepackage{enumitem}
\usepackage{algpseudocode}

\usepackage[T1]{fontenc}

\usepackage[switch,columnwise]{lineno}

\definecolor{green(html/cssgreen)}{rgb}{0.0, 0.5, 0.0}
\definecolor{azure(colorwheel)}{rgb}{0.0, 0.5, 1.0}

\algnewcommand\algorithmicforeach{\textbf{for each}}
\algdef{S}[FOR]{ForEach}[1]{\algorithmicforeach\ #1\ \algorithmicdo}

\def\BibTeX{{\rm B\kern-.05em{\sc i\kern-.025em b}\kern-.08em
		T\kern-.1667em\lower.7ex\hbox{E}\kern-.125emX}}

\begin{document}

\title{Mining User Privacy Concern Topics from App Reviews}

\address[HZNU]{Department of Management Science and Engineering, Hangzhou Normal University, Hangzhou, Zhejiang, P.R.China}
\address[CCNT]{Department of Electrical Engineering and Computer Science, University of Cincinnati, Cincinnati, OH 45221, USA}
\cortext[Corresponding]{Corresponding author}

\author[HZNU]{Jianzhang Zhang}\ead{zjzhang@hznu.edu.cn}
\author[HZNU]{Jinping Hua}\ead{huajinping@stu.hznu.edu.cn}
\author[HZNU]{Yiyang Chen}\ead{yiyang_chen@stu.hznu.edu.cn}
\author[CCNT]{Nan Niu}\ead{nan.niu@uc.edu}
\author[HZNU]{Chuang Liu\corref{Corresponding}}\ead{liuchuang@hznu.edu.cn}



%
\begin{abstract}

\noindent\textbf{\textit{Context}}: As mobile applications (Apps) widely spread over our society and life, various personal information is constantly demanded by Apps in exchange for more intelligent and customized functionality. An increasing number of users are voicing their privacy concerns through app reviews on App stores.


\noindent\textbf{\textit{Objective}}: The main challenge of effectively mining privacy concerns from user reviews lies in the fact that reviews expressing privacy concerns are overridden by a large number of reviews expressing more generic themes and noisy content. In this work, we propose a novel automated approach to overcome that challenge. 

\noindent\textbf{\textit{Method}}: Our approach first employs information retrieval and document embeddings to unsupervisedly extract candidate privacy reviews that are further labeled to prepare the annotation dataset. Then, supervised classifiers are trained to automatically identify privacy reviews. Finally, we design an interpretable topic mining algorithm to detect privacy concern topics contained in the privacy reviews.

\noindent\textbf{\textit{Results}}: Experimental results show that the best performed document embedding achieves an average precision of 96.80\% in the top 100 retrieved candidate privacy reviews. All of the trained privacy review classifiers can achieve an $F_{1}$ value of more than 91\%, outperforming the recent keywords matching baseline with the maximum $F_{1}$ margin being 7.5\%. For detecting privacy concern topics from privacy reviews, our proposed algorithm achieves both better topic coherence and diversity than three strong topic modeling baselines including LDA.




\noindent\textbf{\textit{Conclusion}}: Empirical evaluation results demonstrate the effectiveness of our approach in identifying privacy reviews and detecting user privacy concerns expressed in App reviews.


\end{abstract}

\maketitle

\section{Introduction}

Ubiquitous mobile applications (Apps) significantly facilitate various aspects of daily life ranging from social interactions to entertainment in the era of the digital economy~\cite{wang2021managing}. The intelligent and customized functionalities of Apps are empowered by constantly collecting personal data and analyzing the data in real-time. For example, social Apps usually require users' location to provide local interest groups searching or recommendation services.  Though it is necessary for users to provide some personal information in exchange for more personalized features, some Apps conduct privacy-invading tactics facing the increasingly competitive market~\cite{ebrahimi2022unsupervised, ebrahimi2021mobile}, such as illegally listening~\cite{liao2020measuring}, constantly tracking locations~\cite{li2017static}, and excessively collecting of personal health data~\cite{fan2020empirical}. 

Privacy concerns have drawn wide attention of different stakeholders. Legislation departments around the world successively enact privacy and data protection regulations, e.g., the Health Insurance Portability and Accountability Act (HIPAA) in the U.S.A. and the General Data Protection Regulation (GDPR) in Europe, which have been analyzed by requirements engineering (RE) and security communities to model security requirements and devise compliance detection tools~\cite{kafali2017good,momen2019did,wang2020detecting,amaral2023nlp}. Both Google Play and Apple Store provide privacy labels for Apps in the stores to enhance the transparency of Apps' data practice~\cite{RodriguezJAS23,JainRAS23}. Massive end users are expressing their privacy concerns online, e.g., posting reviews in the App stores~\cite{tao2020identifying,dkabrowski2022analysing}.


App reviews, in the past decade, have become a significant requirements elicitation source, advancing data-driven requirements engineering by the masses and for the masses~\cite{maalej2015toward}. Previous studies have confirmed the significance of user satisfaction for successful App evolution~\cite{li2010user, palomba2015user}. To facilitate App maintenance, existing studies have proposed many methods based on natural language processing (NLP) and machine learning algorithms to classify user reviews according to general categories or dive into specific categories of reviews, e.g., App features and bug issues~\cite{pagano2013user, phong2015mining, maalej2016automatic, shi2017understanding, zhang2019software, gao2021emerging, dkabrowski2022analysing}.


Existing qualitative and quantitative studies have confirmed the addressing of user privacy concerns for the better evolution of software products~\cite{bhatia2018empirical,nguyen2019short,iwaya2023privacy}. For developers, it is necessary to identify user privacy concerns so as to better perform the task of eliciting and implementing privacy requirements~\cite{glinz2007non}. Though significant, privacy requirements, as an essential quality attribute, will emerge a posteriori as the system comes into contact with the hyper-connected setting~\cite{anthonysamy2017privacy}. Moreover, there are only a small percentage of reviews expressing privacy concerns compared with other more general themes and various noise~\cite{tao2020identifying}. The aforementioned \textit{a posteriority} and data sparsity pose two main challenges to effectively handling privacy concerns contained in user reviews. Our previous exploratory study~\cite{zhang2023exploring} has confirmed the significance of identifying and analyzing privacy concerns contained in App reviews to investigate the privacy requirements gap between developers and end users.

To overcome the above challenges, in this work, we propose an automated approach to detect user privacy concerns from App reviews through mining topics from privacy reviews. Specifically, our approach consists of three components. The first component employs semantic similarity computed in document embeddings to retrieve candidate privacy reviews so as to mitigate the data sparsity. The second component is responsible for automatically identifying privacy reviews with supervised text classifiers. For the training dataset of the classifier, positive instances are crafted by manually labeling the retrieved candidate privacy reviews, and negative instances are randomly sampled. The third component further dives into the identified privacy reviews and detects user privacy concerns via our proposed topic mining algorithm, which combines K-means clustering and cluster-based TF-IDF. In summary, our contributions are threefold:
\begin{enumerate}[label=(\arabic*)]
	\item We propose an automated and interpretable approach for detecting user privacy concerns from App reviews. We collected a large dataset of 1,886,838  reviews to empirically evaluate our approach. All of the evaluation materials are publicly available\footnote{https://github.com/zhangjianzhang/PCTD}, including implementation, data, and results.
	\item We build generalized privacy review classifiers based on information retrieval and supervised classification algorithms, achieving a best $F_1$ value of 93.77\% and outperforming the recent keywords matching method with the maximum margin being 7.5\%.
	\item We design a novel unsupervised and interpretable topic mining algorithm to identify user privacy concerns contained in privacy reviews. Our algorithm achieves better topic coherence and diversity than three topic modeling baselines including LDA.
\end{enumerate}

To empirically evaluate the effectiveness of our approach, we formulate the following three research questions (RQs):

\begin{enumerate}[label=(\arabic*)]
	\item \textbf{RQ\textsubscript{1}}: How effective is our document embedding similarity based method in retrieving candidate privacy reviews in order to prepare labeled training set?
	
	\item \textbf{RQ\textsubscript{2}}: How effective are our supervised text classifiers for identifying privacy reviews and which algorithm performs best?
	
	\item \textbf{RQ\textsubscript{3}}: How effective is our topic mining algorithm for detecting user privacy concerns compared with other topic modeling algorithms?
\end{enumerate}


The remainder of this paper is structured as follows: Section II presents the overview of our approach followed by detailed elaborations of each component. Section III presents the experimental settings including data, baselines, evaluation measures, and implementation. Section IV presents the evaluation results to answer the research questions. Threats to validity are discussed in Section V. We outline related work in Section VI and conclude the paper with several potential future extension avenues in Section VII.

\section{Framework}

\begin{figure*}[htbp]
	\centering
	\includegraphics[width=\linewidth]{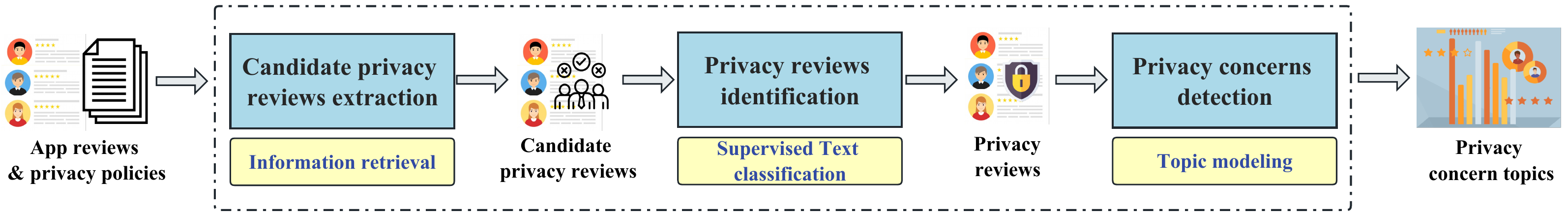}
	\caption{The pipeline of our approach to mining user privacy concern topics from App reviews}
	\label{frame}
\end{figure*}

Figure~\ref{frame} illustrates the pipeline of our approach to mining user privacy concerns from App reviews. The input artifacts include the raw user reviews and privacy policies of a set of Apps. The output is the user privacy concern topics distribution including topic words and representative reviews under each topic. Our approach consists of three components as shown in the dotted line box of Figure~\ref{frame}. 

The functionality and underlying technology of each component are depicted separately in the upper and lower boxes.

\begin{enumerate}[label=(\arabic*)]
	\item The first component is responsible for extracting candidate privacy reviews based on text similarity. The extracted reviews are further assigned binary labels by human experts, 1 for privacy reviews and 0 for privacy irrelevant reviews.
	
	\item The second component employs classification algorithms to learn binary classifiers from the human labeled reviews. The best performed classifier would be used to automatically identify privacy reviews from large scale App reviews.
	
	\item In the third component, we design an interpretable topic modeling algorithm to detect prominent user privacy concerns in the privacy reviews.
\end{enumerate}
We detail the methods employed in each component in the following subsections.

\subsection{Candidate Privacy Reviews Extraction Based on Information Retrieval}



 



App reviews typically contain multiple themes and the quality varies widely ranging from helpful advice to insulting offenses~\cite{pagano2013user}. Supervised text classification has been proved to be effective for review-level requirements information extraction~\cite{maalej2016automatic,kurtanovic2017automatically,jha2019mining}, e.g., functional feature requests and issues as well as non-functional quality attributes. Privacy reviews refer to those expressing potential privacy threats or actual privacy invasions perceived by users. Figure~\ref{fig:review_exp} shows two example privacy reviews and two privacy irrelevant ones.

\begin{figure}
	\centering
	\begin{tcolorbox}[colback=white!20]
		\textbf{Review\textsubscript{1}}: \textit{Why would Instagram need my financial records, health info and audio files among others?} (\textbf{privacy review})
		
		\textbf{Review\textsubscript{2}}: \textit{I closed the location access, but personalized advertisements still appear.} (\textbf{privacy review})
		
		\textbf{Review\textsubscript{3}}: \textit{I just wish it would notify me if I go below a certain dollar amount} (\textbf{privacy irrelevant review})
		
		\textbf{Review\textsubscript{4}}: \textit{Every movement causes the app to crash. Want to open a new drawing Nope, crash.} (\textbf{privacy irrelevant review})
	\end{tcolorbox}
	\caption{Example privacy reviews}
	\label{fig:review_exp}
	\vspace{-1.5em}
\end{figure}

The main difficulty of applying supervised classification to privacy reviews identification lies in preparing training data. Compared to those more common and general App review themes, e.g., feature requests and bug reports, privacy reviews only account for a small percentage. Recent studies show that there are less than 1\% of reviews that express user privacy concerns~\cite{nguyen2019short, tao2020identifying}, leading to the extremely imbalanced ratio of positive and negative privacy review samples. Therefore, it is cost-ineffective and not trivial to directly provide random review samples from the whole reviews set for human annotation when preparing the training data of supervised classifiers. We employ information retrieval to extract candidate privacy reviews for further human annotation. Formally, we define the candidate privacy reviews retrieval procedure as follows:

\textit{\textbf{Document Set}}: $R = \{r_{1}, r_{2}, ..., r_{N}\}$ denotes a set of $N$ user reviews of an App.

\textit{\textbf{Query}}: $q$ is the privacy policy text of the App under consideration.

\textit{\textbf{Relevance Function}}: text similarity is used to quantify the degree of relevance between the query and a document. Specifically, we employ the cosine similarity metric which is defined as follow:
\begin{equation}
CosineSim(q, r) = \frac{\mathbf{q} \cdot \mathbf{r}}{||\mathbf{q}|| \times ||\mathbf{r}||}
\end{equation}
where $\mathbf{q}$ and $\mathbf{r}$ are separately the vector representations of the privacy policy $q$ and a user review $r$. $||\mathbf{q}||$ and $||\mathbf{r}||$ denotes the $L_{2}$ norm of those two vectors respectively. In our approach, we employ pre-trained sentence transformer models to obtain the document embedding vectors:
\begin{equation}
\mathbf{q} = SentenceTransformer(q)
\end{equation}
\begin{equation}
\mathbf{r} = SentenceTransformer(r)
\end{equation}

\textit{\textbf{Document Ranking}}: Based on the established relevance function, we rank the reviews in set $R$ in relation to the given privacy policy as:
\begin{equation}
rank(q, R) = \text{sort}_{r \in R} sim(q, r)
\end{equation}
\( \text{sort}_{d \in D} \) stands for the sorting operation performed on all reviews in set \( R \) according to the cosine similarity score.

\textit{\textbf{Retrieval Output}}: The final retrieval operation yields a subset of the review set in response to the query:
\begin{equation}
\text{output}(q, R) = \{ r \in R \,|\, rank(q, R)\}
\end{equation}

Finally, we select the top $M$ most relevant reviews as the candidate privacy reviews.

The information retrieval technique has been widely used in requirements engineering studies for extracting from various types of artifacts~\cite{niu2013systems,zhao2021natural}. We choose the privacy policy text and document embeddings as the query and ranking criterion other than the domain keywords list and hit count of string matching in previous studies~\cite{tao2020identifying,nema2022analyzing,ebrahimi2022unsupervised}. The rationale for our choices is threefold.

Firstly, the privacy policies cover and further detail those groups of concepts contained in the privacy relevant taxonomies~\cite{anton2004requirements, solove2005taxonomy} that are usually used to build domain keywords list~\cite{nema2022analyzing}. There are four groups of concepts in the taxonomy proposed by \cite{solove2005taxonomy} from the law viewpoint: information collection, information processing, information dissemination, and invasion. Privacy policies are the means to enforce accountability on data protection imposed by legal frameworks. They mainly include four types of statements concerning users' privacy, i.e., how to collect, share, disclose, and retain user data~\cite{caramujo2019rsl}. There are two classes of privacy requirements in the taxonomy proposed by \cite{anton2004requirements}: protection goals and vulnerabilities. This requirements taxonomy is derived from a set of Internet privacy policies. Therefore, the privacy policies can be seen as specific instances of those taxonomies but have more informative statements.

Secondly, compared with abstract taxonomies, there is less language gap between privacy policies and user reviews. Privacy policies are ``notices'' published to end users to inform how their personal data is protected. Besides, privacy policies vary from Apps, which reflects characteristics of different Apps. Hence, the user-oriented and product-specific nature of privacy policies can potentially mitigate the language gap between them and user reviews talking about specific Apps.

Thirdly, the criterion of hit counts of keywords matching does not take into account the meaning and rich vocabulary of reviews, which may omit many relevant reviews. Document similarity, commonly used in information retrieval~\cite{manning2008introduction}, between privacy policies and reviews suffers less from that problem. Moreover, document embeddings trained on large scale corpora can encode the contextual semantics of text~\cite{otter2020survey}, which can further bridge the vocabulary gap between privacy policies and user reviews.


\subsection{Privacy Reviews Identification Based on Text Classification}






Given the returned ranking list, the top $M$ reviews are selected as candidate privacy reviews, and random samples from the list after top $M$ reviews are used to prepare candidate privacy irrelevant reviews. The second and third authors are responsible for annotating the above positive and negative candidate reviews with label $1$ denoting privacy reviews and label $0$ denoting privacy irrelevant ones. Both of the annotators are software engineering postgraduates with practical App development and requirements analysis experience. To ensure the annotation quality, the annotators participate in a 30 minutes training session and are informed of the annotation guideline via illustrative review examples. After training, two annotators independently label the review candidates. For those reviews with inconsistent labels, the first author discusses with the two annotators to determine the final label. Consequently, we obtain a set of labeled reviews.

We model the task of privacy reviews identification as a binary classification problem which can be solved by training supervised text classifiers with the labeled data. Formally, let $C = \{c_{1} = 0,\ c_{2} = 1\}$ denote the class set, the mapping relations from reviews to classes $f: R \rightarrow C$ are learned through training the classification algorithms on the labeled review set $T = \{(r_{1}, c_{1}), (r_{1}, c_{1}), ..., (r_{n}, c_{n})\}$. For a newly coming review instance $r$, its class label $c = f(r)$ is used to determine if it is a privacy review. The best performed classifier is then used to identify privacy reviews from the raw unlabeled App review dataset.

\subsection{Privacy Concerns Detection Based on Topic Modeling}

We design our privacy concerns detection algorithm by adapting a neural topic modeling framework named BERTopic~\cite{grootendorst2022bertopic} which has been widely applied to document collections of various domains, e.g., construction~\cite{garcia2022machine}, climate~\cite{falkenberg2022growing}, and education~\cite{zankadi2023identifying}. The BERTopic framework is composed of a cascade of three modules, i.e., document vectorization, document clustering, and topic words extraction. The Transformer-based document embeddings are employed to convert the raw text to numerical vectors. Hierarchical density-based spatial clustering (HDBSCAN) is used to group those documents into different clusters each of which will be assigned one topic. Each topic is represented by a ranked word list where the words are extracted from the documents in the corresponding cluster based on the importance of a word to that cluster.

Our adaptions of BERTopic for detecting privacy concerns mainly focus on two aspects. 

Firstly, we determine the document embedding model used in the document vectorization module via experiments. Specifically, we use the best performed document embedding in the candidate privacy reviews extraction component of our approach. Because information retrieval acts as an important task for evaluating a document embedding's capability of encoding meaningful semantic information~\cite{li2022brief}. 

Secondly, we replace the HDBSCAN with K-means in the text clustering module. The rationale of this substitute is twofold. On the one hand, the clusters obtained by performing K-means on the document embeddings generated by various pre-trained models can be interpreted as topics, similar to those discovered by LDA~\cite{zhao2021topic}. On the other hand, K-means has been proved to be effective in App reviews analysis by previous studies~\cite{carreno2013analysis,nema2022analyzing}. The number of App reviews is large and increasing rapidly leading to the need for a more computationally effective algorithm. The hierarchical clustering algorithms have a complexity that is at least quadratic in the number of documents compared to the linear complexity of K-means~\cite{manning2008introduction}.

We formulate our adaptions of BERTopic for detecting privacy concern topics in Algorithm~\ref{algo:topic}.

\begin{algorithm}[h]
	\caption{Privacy Concern Topics Detection (PCTD)}
	\begin{algorithmic}[1]
		\Procedure{PCTD}{$R_{p}$, $K$}
		\State $\{\vec{r_{1}},\vec{r_{2}}, ...,\vec{r_{N}}\}$ $\leftarrow$ \textit{DocEmbeddingVectorization}($\mathcal{R}_{p}$)
		\State $\{\vec{x_{1}},\vec{x_{2}}, ...,\vec{x_{N}}\}$ $\leftarrow$ \textit{Normalization}($\{\vec{r_{1}},\vec{r_{2}}, ...,\vec{r_{N}}\}$)
		\State $\{\vec{s_{1}},\vec{s_{2}}, ...,\vec{s_{K}}\}$ $\leftarrow$ \textit{SelectSeeds}($\{\vec{x_{1}},\vec{x_{2}}, ...,\vec{x_{N}}\}$, $K$)
		\For{k = 1 to K}
		\State $C_k \leftarrow \{\}$ \Comment{Initialize clusters}
		\State $\vec{c_{k}} \leftarrow \vec{s_{k}}$ \Comment{Initialize cluster centroids}
		\EndFor
		\While{stopping criterion has not been met}
		\For{each $\vec{x_{i}} \in \{\vec{x_{1}},\vec{x_{2}}, ...,\vec{x_{N}}\}$}
		\State $k_{^*} = \arg\min_{k \in \{1,\dots,K\}} \|\vec{x_{i}} - \vec{c_{k}}\|^2$ \Comment{Assign $\vec{r_{i}}$ to cluster $k^*$ of which the centroid is closest to $\vec{r_{i}}$}
		\State $C_{k_{^*}} \leftarrow C_{k_{^*}} \cup r_{i}$ \Comment{Update clusters}
		\EndFor
		\For{k = 1 to K}
		\State $\vec{c_{k}} = \frac{1}{|C_k|} \sum_{\vec{x_{i}} \in C_{k}} \vec{x_{i}}$
		\Comment{Update cluster centroids}
		\EndFor
		\EndWhile
		\For{each single word $w \in \mathcal{R}_{p}$}
		\State $w_{k} \leftarrow \textit{ClusterBasedTF-IDF}(w)$ \Comment{Calculate the importance of a word $w$ to a cluster $C_{k}$}
		\EndFor
		\For{k = 1 to K}
		\State $T_{k} = [t_{1}, t_{2}, ...,]_{t \in C_{k}} \leftarrow \text{Sort}_{w \in C_{k}} \{w\}$ \Comment{Sort the words in each cluster by their importance}
		\EndFor
		\State \Return $\{(C_{1}, T_{1}), (C_{2}, T_{2}), ..., (C_{K}, T_{K})\}$
		
		\EndProcedure
		
	\end{algorithmic}
	\label{algo:topic}
\end{algorithm}

The input of Algorithm~\ref{algo:topic} consists of a list of $N$ privacy reviews list $\mathcal{R}_{p} = \{r_{1},r_{2}, ...,r_{N}\}$ and the number of topics $K$. The output includes review clusters along with an ordered list of words by their importance to the corresponding cluster. Each cluster $C_{k}$ is treated as a privacy concern topic and the top ranked words in $T_{k}$ are used to represent the topic meaning of that cluster. We further elaborate the main body of Algorithm~\ref{algo:topic} as follows:

\textbf{Lines 2-3 (Vectorization and Normalization)}: Document embedding is used to convert the privacy reviews to numeric vectors. $L_{2}$ normalization is applied to the vectors:
\begin{equation}
\vec{x_{i}} = \frac{\vec{r_{i}}}{\|\vec{r_{i}}\|}
\label{eq:norm} 
\end{equation}
The normalization in equation~\eqref{eq:norm} is also referred to as length-normalization which aims to eliminate the impact of document vector length on the document distance calculation~\cite{manning2008introduction}. After applying the length-normalization, Euclidean distance used in K-means (see Line 9 in Algorithm~\ref{algo:topic}) therefore becomes an effective proxy for the cosine distance (i.e., $1-ConsineSim$) as shown below:

\begin{align}
Euclidean\ Distance(\vec{x_{i}}, \vec{x_{j}})^2 &= \|\vec{x_{i}} - \vec{x_{j}}\|^2\\
&= \sum_{n}(x_{in}-x_{jn})^2\\
&= 2\times(1-\vec{x_{i}} \cdot \vec{x_{j}})\\
&=2\times CosineDistance(\vec{x_{i}}, \vec{x_{j}})
\end{align}
where $n$ is the dimension of embedding vectors of $\vec{x_{i}}$, $\vec{x_{j}}$.

\textbf{Lines 4 - 17 (K-means Clustering)}: For overcoming the curse of dimensionality~\cite{pandove2018systematic}, the reducing dimensionality method UMAP~\cite{mcinnes2018umap} is employed, which has been shown to preserve more of the local and global features of high-dimensional data in lower projected dimensions. The initial cluster centroids are determined by the K-means++~\cite{arthur2007k}, which selects initial cluster centroids using sampling based on an empirical probability distribution of the data points' contribution to the overall inertia. 

\textbf{Lines 18 - 23 (Topic Words Selection)}: For each cluster of privacy reviews $C_{k}$, the importance of a word $w$ to that cluster $C_{k}$ is measured with cluster-based TF-IDF, which is defined as follow:
\begin{equation}
w_{k} = TF_{w,c} \cdot \log(1+\frac{A}{TF_{w}})
\label{eq:c-tf-ifd}
\end{equation}
where $TF_{w,c}$ is the frequency of $w$ in the cluster $C_{k}$, $TF_{w}$ is the total frequency of word $w$ in the corpus, and $A$ denotes the average number of words per cluster. After sorting the words by their cluster-based TF-IDF for each cluster $C_{k}$, an ordered word list $T_{k}$ is obtained. The top ranked words in $T_{k}$ are finally selected as topic words for the cluster $C_{k}$.

\section{Experimental Settings}


\subsection{Data Collection}

\textbf{Target Apps Selection}: The target Apps are selected based on the popularity typically indicated by the number of downloads~\cite{martin2015app}. The underlying rationale of this choice is that popular Apps have a large user base which ensures the variety of App users and the sufficient number of reviews for performing App-Store inspired requirements elicitation~\cite{maalej2015toward, ferrari2023strategies}. As popular Apps are commonly used, analyzing users' privacy concerns about them provides a helpful understanding what is the current benchmark on users' privacy requirements in the market. To determine the popularity of Apps, we refer to the rank lists of English speaking markets on App Annie following the practice in previous studies~\cite{hassan2018studying, assi2021featcompare}. We further check the number of reviews of Apps in the rank lists and remove those newly launched Apps having a surging growth trend but limited user reviews. Finally, we select 11 Apps across 4 categories as shown in Table~\ref{tbl:reviews}.

\textbf{User Reviews Collection}: We collect the user reviews of the target Apps from the App analytics platform QiMai\footnote{https://www.qimai.cn/} which is also used in previous App user reviews study~\cite{wu2021identifying}. User reviews of the target Apps on Apple Store are gathered as Google Play has an acquisition constraint on the number of user reviews~\cite{pagano2013user,jha2019mining}. The posting time spans from 2020-01-01 to 2023-03-20. The market regions include the U.S.A., Canada, the UK, and Australia. In total, we collect 1,886,838 reviews 

\textbf{Privacy Policies Collection}: We collect each App's privacy policy document from its official website and remove the ``\textit{generic}'' statements that do not involve App privacy data practice~\cite{caramujo2019rsl}, e.g., \textit{how to contact with App vendors} and \textit{how to notify policy change}.

\begin{table}[!ht]
	\centering
	\caption{App Reviews Summary Statistics}
	\begin{tabular}{c|c|c}
		\toprule
		\textbf{Category} & \textbf{App} & \textbf{\# of reviews}  \\
		\midrule
		\multirow{2}{*}{Social Networking} & Facebook &    262,341\\ 
		\cline{2-3}
		& Twitter &  11,375\\
		\hline
		
		\multirow{4}{*}{Multimedia Sharing} & Instagram &    255,899\\
		\cline{2-3}
		& YouTube &   252,603\\
		\cline{2-3}
		& TikTok &   544,236\\
		\cline{2-3}
		& Spotify &   231,995\\
		\cline{2-3}
		
		\hline
		\multirow{2}{*}{Instant Messaging} & Snapchat &    160,614\\
		\cline{2-3}
		& WhatsApp &  44,781 \\
		
		\hline
		\multirow{3}{*}{Community Platforms} & Reddit &   38,986 \\
		\cline{2-3}
		& Discord &   45,943\\
		\cline{2-3}
		& Pinterest &   38,065\\
		\hline
		\textbf{Total} & \multicolumn{2}{c}{1,886,838}\\
		\bottomrule
	\end{tabular}
	\label{tbl:reviews}
\end{table}


\subsection{Baselines}




\textbf{Document Embeddings}: For converting privacy policies and reviews to numeric vectors, we employ the SentenceTransformers framework~\cite{reimers2019}, which offers a collection of pre-trained text embedding models tuned for various tasks, e.g., textual similarity computation, semantic search, and paraphrase mining. Specifically, we compare 4 embedding models considering both the   quality and speed\footnote{https://www.sbert.net/docs/pretrained\_models.html\#sentence-embedding-models/}. The brief descriptions of the selected 4 embedding models are listed as follows:

\begin{itemize}
	\item \textit{all-mpnet/all-MiniLM}: These models are trained on more than 1 billion  training pairs and designed for the general purpose (\textit{all}). 
	
	\item \textit{multi-qa-mpnet/multi-qa-MiniLM}: These models are trained on 215M question-answer pairs from various sources (\textit{multi-qa}) and domains and specifically trained for semantic search.
\end{itemize}

Each model name is composed of two parts. The first part denotes the design purpose. In the second part, \textit{mpnet} means the model provides the best quality on the extensive evaluations while \textit{MiniLM} means the model is faster and still offers good quality.

\textbf{Classification Algorithms}: For building the privacy review classifiers, we compare three categories of classification algorithms that are described as follows:

\begin{itemize}
	\item \textit{Logistic Regression/Linear SVM}: These classic linear models are widely used as baselines in the text classification task. 
	
	\item \textit{Random Forest/Gradient Boosting}: These ensemble algorithms are strong competitors in various data competitions.
	
	\item \textit{fastText/BERT}: These deep neural network (DNN) models have achieved new state of the arts in many text classification benchmarks.
\end{itemize}

For fastText and BERT, we use their pre-trained word embeddings\footnote{https://fasttext.cc/docs/en/english-vectors.html}\footnote{https://huggingface.co/bert-base-uncased} as input feature vectors. In particular, BERT is attached with a bi-directional GRU layer whose parameters are fine-tuned with the training data.

\textbf{Privacy Reviews Identification Baseline}: The baseline is a recently proposed semi-supervised privacy reviews identification method~\cite{ebrahimi2022unsupervised}. Starting with an initial set of general privacy related keywords, the baseline identifies privacy reviews and new keywords in an iterative manner, where human efforts are needed to judge the newly identified keywords in each iteration.

\textbf{Topic Modeling Baselines}: We compare our adapted BERTopic with the traditional bag-of-words topic model LDA~\cite{blei2003latent} and the recent neural topic model ETM~\cite{dieng2020topic} in addition to the original BERTopic.

\subsection{Evaluation Measures}

\textbf{Information Retrieval}: For evaluating the performance of candidate privacy reviews retrieval, we employ the measures of $F_{1}@k$ and Average Precision (AP)~\cite{manning2008introduction}. $F1@k$ is the harmonic mean of $precision@k$ and $recall@k$. $precision@k$ is the fraction of relevant items among the retrieved $k$ items, i.e., 
\begin{equation}
precision@k = \frac{\#\ of\ retrieved\ relevant\ items}{k}
\end{equation}
$recall@k$ is the fraction of relevant items that are retrieved, i.e., 
\begin{equation}
recall@k = \frac{\#\ of\ retrieved\ relevant\ items}{\#\ of\ relevant\ items}
\end{equation}

AP is defined as follow:
\begin{equation}
AP = \frac{\sum_{r =1}^{k} P(r) \cdot rel(r)}{\#\ of\ relevant\ items}
\label{eq:ap}
\end{equation}
where $P(r)$ is the precision when the result list is treated as containing only its first $r$ items, and $rel(r)$ equals 1 if the $r_{th}$ item is relevant and 0 otherwise.

\textbf{Text Classification}: For evaluating the performance of privacy reviews classification, we adopt the commonly used precision, recall, and $F_1$ score.

\textbf{Topic Modeling}: For quantitatively evaluating the performance of topic modeling algorithms, we adopt the topic coherence measures which indicate the interpretability of the output topics~\cite{blei2003latent}. Specifically, we employ the coherence value $C_{V}$ proposed by~\cite{roder2015exploring}, which outperforms existing topic coherence measures with respect to correlation to human ratings on the systematic large scale evaluation. $C_{V}$ combines the indirect cosine measure with the normalized pointwise mutual information and the boolean sliding window. Besides topic coherence, topic diversity~\cite{dieng2020topic} is also computed to measure the redundancy of generated topics, which is the percentage of unique topic words for all topics.

\subsection{Implementation and Parameters}

We implement our approach and the baselines with Python and some well-known open source packages. In the candidate privacy reviews extraction component, we use the pre-trained document embeddings in SentenceTransformer\footnote{https://www.sbert.net/index.html}. In the privacy reviews identification component, the classification algorithms and K-means algorithm are implemented by using scikit-learn\footnote{https://scikit-learn.org/}. We implement the BERT classifier with the Transformers package \footnote{https://github.com/huggingface/transformers} and the implementation of fastText is provided by Facebook\footnote{https://fasttext.cc/docs/en/python-module.html}. In the privacy concerns detection component, the implementation of BERTopic\footnote{https://maartengr.github.io/BERTopic/index.html} and ETM\footnote{https://github.com/adjidieng/ETM} are provided by the original papers. The LDA is implemented with Gensim. 

For a fair comparison, we do not tune the hyper parameters of the compared algorithms and use the default parameters in those packages.




%
%

\section{Experiment Results Analysis}

In this section, we present and analyze the evaluation results of each component of our approach to answer the research questions.

\subsection{Candidate Privacy Reviews Extraction}\label{sec:ir}


For evaluation purpose, we first establish a ground truth with a subset of reviews of Facebook due to its huge user base and large review size. Specifically, the Facebook's privacy policy is the query, and a subset of its 14,122 reviews act as the documents set. We use the document embedding \textit{multi-qa-MiniLM} to compute the cosine similarity between the query and each review. The reviews are then sorted by the similarity scores in descending order. The first 500 reviews are manually labeled with 1 (privacy reviews) or 0 (privacy irrelevant reviews). The resulting ground truth contains 255 privacy reviews and 245 irrelevant ones.

The other three models are further evaluated on this ground truth. In other words, the 500 reviews are re-sorted by the computed similarity scores based on those three document embedding models. The better a model performs, the more privacy reviews will appear at the top of the sorted review list. As more top ranked reviews are included in the final retrieval result, both the numbers of privacy reviews and irrelevant ones increase, which means that the precision decreases while the recall increases. For trade-off consideration, we compute $F1@k$, where $k$ is the size of the retrieval result. 

\begin{figure}[htbp]
	\centering
	\includegraphics[width=\linewidth]{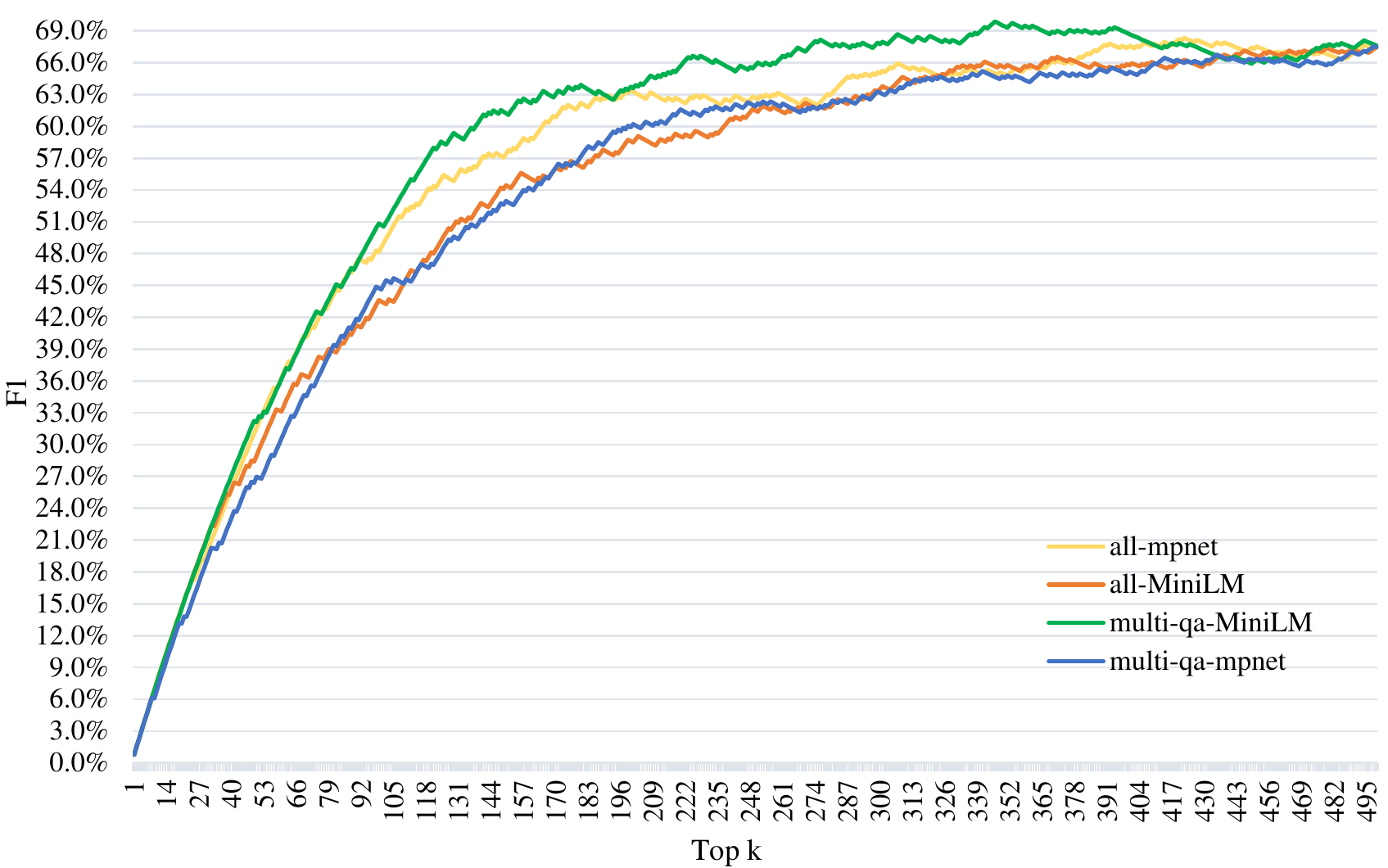}
	\caption{Retrieval Performance of Four Document Embeddings}
	\label{fig:ftopk}
	\vspace{-1em}
\end{figure}

Figure \ref{fig:ftopk} shows the $F1@k$ achieved by four different document embedding models. As $k$ increases, the $F1$ scores of all models firstly increase rapidly and then become stable. In most cases (i.e., different $k$ values), the \textit{multi-qaMiniLM} model performs better than others. Both \textit{all-MiniLM} and \textit{multi-qa-mpnet} achieve relatively lower $F1$. The \textit{all-mpnet} model performs better than \textit{all-MiniLM} and \textit{multi-qa-mpnet} when $k < 235$. The F1 scores of all models stabilize at about 67\% as $k$ increases near 500. The highest $F1$ of 69.88\% is achieved by \textit{multi-qaMiniLM} at $k = 346$ and the corresponding recall is 82.35\%. Table \ref{tbl:retrieval_ap} shows the average precision of the top 100 retrieval results returned by different models.

\begin{table}[!ht]
	\centering
	\caption{Average Precision of Different Models (Top 100)}
	\begin{tabular}{c|c}
		\toprule
		\textbf{Models} & \textbf{AP}\\
		\hline
		all-mpnet & 94.83\%\\
		\hline
		all-MiniLM & 93.86\%\\
		\hline
		multi-qa-MiniLM & \textbf{96.80\%}\\
		\hline
		multi-qa-mpnet & 88.87\%\\
		\bottomrule
	\end{tabular}
	\label{tbl:retrieval_ap}
\end{table}

As shown in Table~\ref{tbl:retrieval_ap}, \textit{multi-qa-MiniLM} achieves the best AP of 96.80\% with a margin of 7.93\% compared with the lowest AP achieved by \textit{multi-qa-mpnet}. 

We finally select \textit{multi-qa-MiniLM} to convert the privacy policies and reviews to vectors of which the cosine similarity is computed for retrieving candidate privacy reviews. Specifically, we used the 6 Apps from the Social Networking and Multimedia Sharing categories in Table~\ref{tbl:reviews} as they involve a large and diverse user base. The reviews of each App are sorted by similarity scores in descending order. The top 500 reviews of each App are extracted as candidate privacy reviews which are further annotated to prepare classification training data.


\textbf{Answer to RQ\textsubscript{1}}: \textit{Overall, the four comparative document embeddings achieve an $F_{1}$ of 67\% evaluated on a test set of 500 reviews. \textit{multi-qa-MiniLM} achieves the best performance with an $F_{1}@k$ of 69.88\% and an average precision of 96.80\% at $k=100$.}

\subsection{Privacy Reviews Identification}
Two software engineering postgraduates are responsible for independently annotating the extracted candidate privacy reviews. One postgraduate firstly annotates the top 500 extracted reviews of each App. Another postgraduate then annotates the same review set without knowing the labels. They are allowed to skip the uncertain ones. Finally, they annotate 2,769 reviews in total together. The inter-annotator agreement, computed using Cohen's kappa coefficient, is 0.88, indicating ``almost perfect agreement"~\cite{landis1977measurement}. The first author further discusses with them to resolve the disagreement. The final labeled data set is unbalanced including 1,040 privacy related reviews and 1,729 negative samples. we further employ the random undersampling to achieve a balanced class distribution following previous practice~\cite{villarroel2016release,kurtanovic2017automatically}. To evaluate various classification algorithms, we randomly partition the labeled reviews into training and testing sets at an 8:2 ratio.

Table~\ref{tbl:cls_acc} reports the performance of different classifiers in identifying privacy reviews. For precision, all classifiers have a score exceeding 93\%. \textit{BERT} and two ensemble algorithms obtain a precision higher than 95\%. The highest precision of 97.41\% is achieved by \textit{Gradient Boosting}. For recall, only \textit{Gradient Boosting} and two DNN algorithms obtain scores exceeding 90\% with the same value of 90.38\%. \textit{Random Forest} has the lowest recall of 88.94\%. When taking the precision and recall into consideration together, all classifiers achieve an F1 score higher than 91\%. Gradient Boosting achieves the highest F1 score of 93.77\%. Overall, the two linear algorithms, i.e., \textit{Logistic Regression} and \textit{linear SVM} perform worse than the ensemble and DNN algorithms. \textit{Gradient Boosting} performs consistently better than the others with the highest precision, recall, and F1 score.

\begin{table}[!ht]
	\centering
	\caption{Classification Performance of Different Classifiers}
	\begin{tabular}{c|c|c|c}
		\toprule
		\textbf{Algorithms} & \textbf{Precision} & \textbf{Recall} & \textbf{F1}  \\
		\midrule
		Logistic Regression & 93.00\% & 89.42\% & 91.18\%  \\ \hline
		linear SVM & 93.47\% & 89.42\% & 91.40\%  \\ \hline
		Random Forest & 95.36\% & 88.94\% & 92.04\%  \\ \hline
		Gradient Boosting & \textbf{97.41\%} & \textbf{90.38\%} & \textbf{93.77\%}  \\ \hline
		fastText & 94.00\% & \textbf{90.38\%} & 92.16\%  \\ \hline
		BERT & 95.43\%  & \textbf{90.38\%}  & 92.84\%  \\
		\bottomrule
	\end{tabular}
	\label{tbl:cls_acc}
\end{table}

\begin{figure}[htbp]
	\centering
	\includegraphics[width=\linewidth]{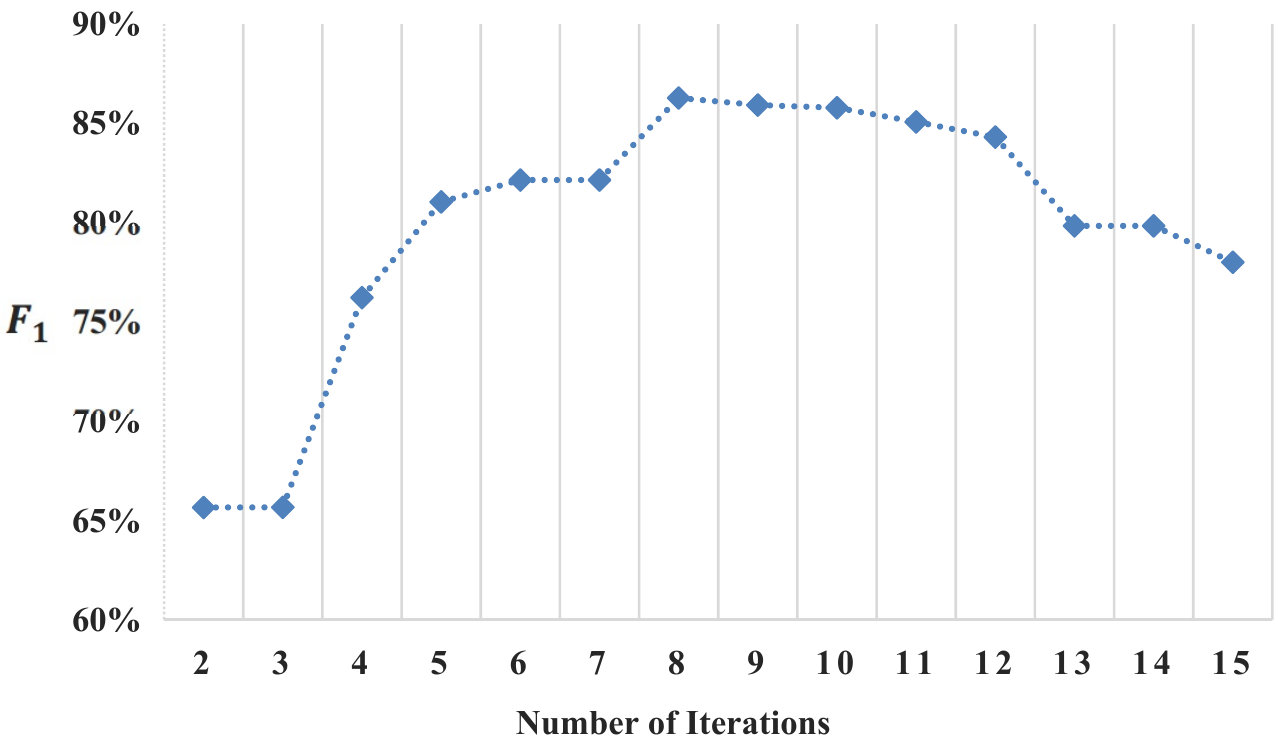}
	\caption{Performance of the Privacy Reviews Identification Baseline}
	\label{fig:cls-compare}
	\vspace{-1em}
\end{figure}

On the same testing set, we evaluate the performance of the privacy reviews identification baseline. Privacy keywords are iteratively selected from the same training set. Figure~\ref{fig:cls-compare} depicts the evaluation results. The horizontal axis denotes the number of iterations and the vertical axis denotes $F1$. The $F_{1}$ score reaches its peak of 86.27\% at the eighth iteration. It can be observed from Figure~\ref{fig:cls-compare} that the $F_{1}$ score exhibits an initial rise followed by a decline as the number of iterations increases. The ``rise-then-fall'' trend observed in the F1 score can be attributed to the bootstrapping nature of the baseline method, which suffers from the problem of ``semantic drift''~\cite{mintz2009distant,wang2017aspect}.

We perform quantitative error analysis of the classification results of the best performed algorithm \textit{Gradient Boosting}. Table~\ref{tbl:confusion} shows the confusion matrix. The number of false negative instances is far more than the number of false positive ones. This observation accords with the result in Table~\ref{tbl:cls_acc} that all classifiers achieve a higher value on precision compared with recall. 

\begin{table}
	\caption{The Confusion Matrix of Gradient Boosting}
	\begin{tabular}{l|l|c|c|c}
		\multicolumn{2}{c}{}&\multicolumn{2}{c}{Prediction}&\\
		\cline{3-4}
		\multicolumn{2}{c|}{}&Privacy &Irrelevant&\multicolumn{1}{c}{Total}\\
		\cline{2-4}
		\multirow{2}{*}{Truth}& Privacy & $203$ & $5$ & $208$\\
		\cline{2-4}
		& Irrelevant & $20$ & $188$ & $208$\\
		\cline{2-4}
		\multicolumn{1}{c}{} & \multicolumn{1}{c}{Total} & \multicolumn{1}{c}{$223$} & \multicolumn{1}{c}{$193$} & \multicolumn{1}{c}{$416$}\\
	\end{tabular}
	\label{tbl:confusion}
\end{table}

We further qualitatively investigate the content of those error instances via human content analysis~\cite{neuendorf2017content}. Most false negative (FN) reviews mix privacy concerns with too much irrelevant content. This is consistent with the previous empirical finding~\cite{pagano2013user} that App reviews typically contain diverse topics. False positive (FP) instances usually contain words that frequently appear in the context of expressing privacy concerns, e.g., \textit{sell}, \textit{information}, but their content do not express any privacy concern in fact. Figure \ref{fig:err_example} shows two error example reviews belonging to these two types.

\begin{figure}
	\centering
	\begin{tcolorbox}[colback=white!20]
		\textbf{FN}: \textit{Creepy Tracking and Censorship of wrong think. YouTube was good once upon a time. Now they censor opinions they dont like. Track your usage and manipulate your search results. They are by extension a malicious foreign actor and should be handled with caution}
		
		\textbf{FP}: \textit{Facebook is done. Too much censorship, cant express yourself freely anymore. Fact checkers is everywhere with info theyve got from unreliable sources. Facebook groups wont allow to even make a post anymore unless its a for sell post overall everything has went downhill.}
		
	\end{tcolorbox}
	\caption{False Negative and False Positive Classification Examples}
	\label{fig:err_example}
\end{figure} 

\textbf{Answer to RQ\textsubscript{2}}: \textit{Our text classification based methods outperform the semi-supervised keywords matching baseline with the largest margin of 7.5\% on $F_{1}$. The best performed classification algorithm is Gradient Boosting which performs consistently well on precision (97.41\%), recall (90.38\%), and $F_{1}$ (93.77\%).}

\subsection{Privacy Concerns Detection}

%
%
%
%
%
%

\begin{figure}[htbp]
	\centering
	\includegraphics[width=0.56\linewidth]{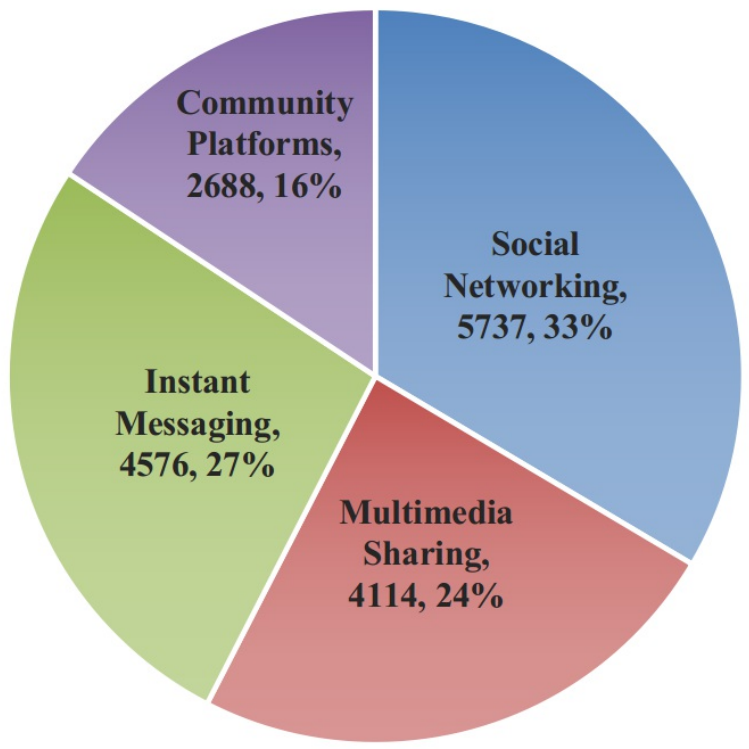}
	\caption{Privacy Reviews for Topic Modeling Input}
	\label{fig:pie}
	\vspace{-1em}
\end{figure}

We apply the trained Gradient Boosting classifier to extract privacy reviews of all Apps listed in Table~\ref{tbl:reviews}. The privacy reviews of Multimedia Sharing Apps are dominant in the identification results due to the large number of raw reviews as shown in Table~\ref{tbl:reviews}. To more effectively evaluate various topic modeling techniques, we construct a data source balanced dataset which is a common practice in the benchmark datasets for topic modeling~\cite{roder2015exploring}. Specifically, we built an App category balanced privacy reviews dataset by including the privacy reviews of Instagram in the Multimedia Sharing category and the privacy reviews of all apps in the other three categories. Figure~\ref{fig:pie} shows the distribution of privacy reviews used for topic modeling.

\begin{figure}[htbp]
	\centering
	\includegraphics[width=\linewidth]{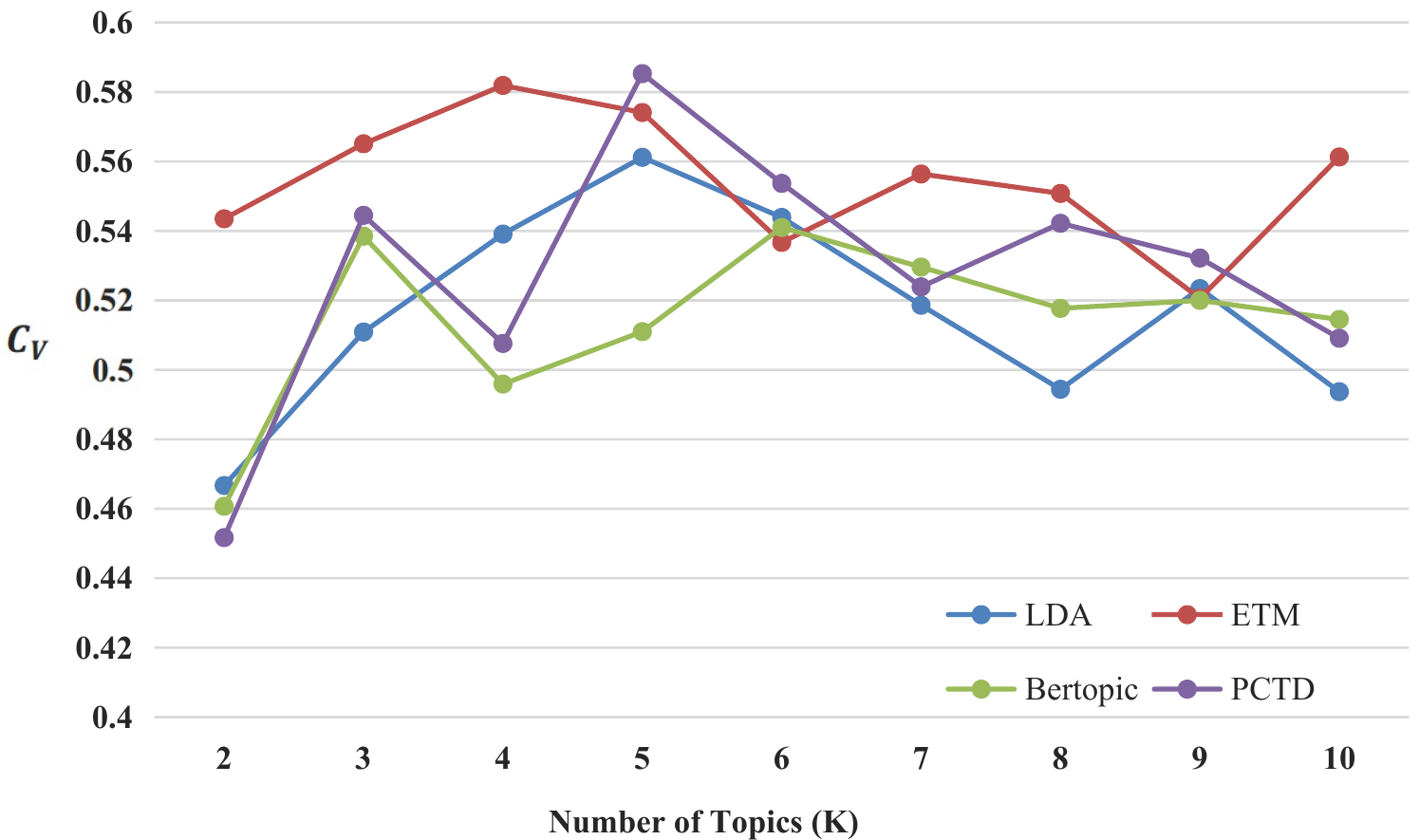}
	\caption{Topic Coherence $C_{V}$ for Different Algorithms with Varying Number of Topics $K$}
	\label{fig:cv}
	\vspace{-1em}
\end{figure}

We select the \textit{multi-qa-MiniLM} embedding to vectorize privacy reviews in our PCTD algorithm, which exhibits best candidate privacy reviews retrieval performance thus demonstrating its superiority in encoding the semantics of privacy reviews over other embeddings. The measures of topic coherence and topic diversity are computed under each number of topics $K$ ranging from 2 to 10 with a step size 1. Figure~\ref{fig:cv} shows the $C_{V}$ values of four topic modeling algorithms along with different numbers of topics.

Figure~\ref{fig:cv} illustrates that the $C_{V}$ values of all algorithms first rise and subsequently decrease as $K$ varies, peaking when $K$ lies between 4 and 6. Among the best $C_{V}$ values attained by each algorithm, our $PCTD$ achieves the highest of 0.5853 at $ K = 5 $. The ETM closely follows with its best $C_{V}$ value of 0.5819 at $ K = 4 $. The LDA reaches its optimal CV value of 0.5612 at $ K = 5 $, while the original BERTopic records the lowest value of 0.5612 at $ K = 5 $.

We further compute the topic diversity of each topic modeling algorithm for their optimal numbers of topics $K$ in terms of $C_{V}$. For each topic, the top 10 words belonging to that topic are selected as topic words. We determine the ranking of words under each topic according to the probabilities of word document distribution (for LDA and ETM) and the cluster-based TF-IDF defined in formula~\eqref{eq:c-tf-ifd} (for BERTopic and our $PCTD$). Table~\ref{tbl:diversity} shows the topic diversity values of each topic modeling algorithm with the optimal number of topics $K$.

\begin{table}[!ht]
	\centering
	\caption{Topic Diversity of Each Topic Modeling Algorithm}
	\begin{tabular}{c|c|c|c}
		\toprule
		  \textbf{LDA} & \textbf{ETM} & \textbf{BERTopic} & \textbf{PCTD} \\
		\midrule
		0.7500 & 0.8250 & 0.7083 & \textbf{0.8800}\\
		\bottomrule
	\end{tabular}
	\label{tbl:diversity}
\end{table}

As shown in Table~\ref{tbl:diversity}, our $PCTD$ achieves the highest topic diversity value of 0.88. The higher topic diversity value means that there are more various topic words to represent the topics, which facilitates human analysts' interpretation and naming of detected topics. Table~\ref{tbl:topics} displays the top 10 topic words for each topic detected by our $PCTD$.

\begin{table}
	\caption{Privacy Concern Topics}
	\begin{tabular}{c|m{5.4cm}}
		\toprule
		\textbf{Topics} & \textbf{Topic Words}\\
		\midrule
		\textbf{Data Selling} & datum, sell, privacy, data, steal, private, social, business, information, money \\ \hline
		\textbf{Account Security} & hack, hacker, account, password, login, ban, access, security, email, recover \\ \hline
		\textbf{Data Security} & reinstall, error, log, reset, restart, connection, connect, login, datum, refresh \\ \hline
		\textbf{Tracking} & privacy, profile, issue, message, delete, problem, account, social, notification, private \\ \hline
		\textbf{Privacy Control} & privacy, private, security, secure, social, information, access, hide, trust, personal \\
		\bottomrule
	\end{tabular}
	\label{tbl:topics}
\end{table}

\begin{figure}[h]
	\centering
	\includegraphics[width=\linewidth]{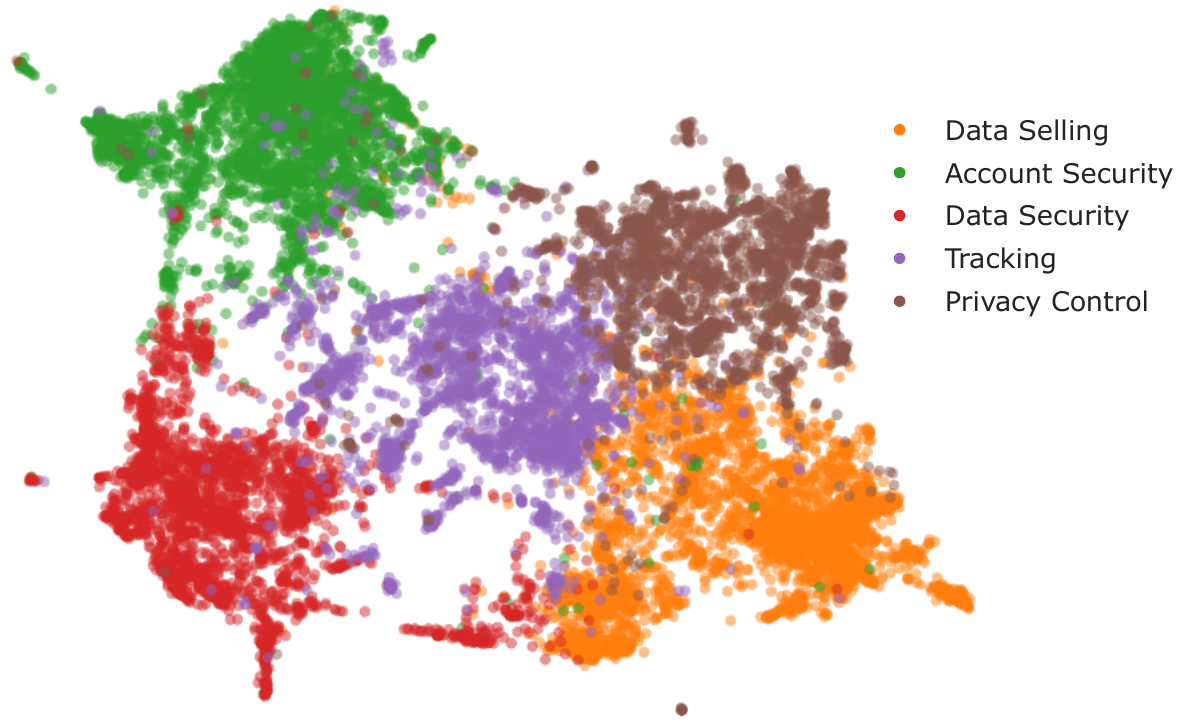}
	\caption{Privacy Concern Topics Visualization}
	\label{fig:visual}
	\vspace{-1em}
\end{figure}

In Table~\ref{tbl:topics}, each topic is named by investigating the topic words and the representative reviews belonging to that topic. As each cluster found by our $PCTD$ denotes one topic, we select the top 10 reviews from each cluster to serve as the representative reviews based on the cosine similarity between the review embedding and the corresponding cluster centroid. Figure~\ref{fig:visual} illustrates the distribution of privacy clusters on a 2-D plane. It is evident that the reviews under different topics form distinct and cohesive regions, which demonstrates the effectiveness of our $PCTD$ in differentiating semantically related reviews. We attribute those outliers of each cluster to the privacy reviews identification errors and the noisy nature of App reviews~\cite{pagano2013user}.

As can be seen in Table~\ref{tbl:topics} and Figure~\ref{fig:visual}, five major privacy concerns are detected from the privacy reviews including:
\begin{itemize}
	\item \verb|Data selling|: App providers sell users' data to others for profits.
	
	\item \verb|Account security|: Users' accounts are not secure enough to defend the potential hacking.
	
	\item \verb|Data security|: There are sometimes bug issues when users try to transfer and restore their personal data.
	
	\item \verb|Tracking|: Various permissions and personal information are constantly tracked by Apps even if not necessary, e.g., messages, notifications, camera, microphone voice etc.
	
	\item \verb|Privacy control|: Users complain about the current privacy settings and request new features to enhance privacy control.
\end{itemize}
Figure~\ref{fig:privacy_concern_exp} presents the representative review examples of each privacy concern.

\begin{figure}[h]
	\centering
	\begin{tcolorbox}[colback=white!20]
		\textbf{Data Selling}: \textit{An application collecting data about your interests to sell it to merchants and advertisers.}
		
		\textbf{Account Security}: \textit{I'm unable to sign in or create a new account cause my old one is hacked with my email and phone number.}	

		\textbf{Data Security}: \textit{Broken App. I try to open, it will automatically shutdowns.I re installed the app and try to restore data from iCloud, I lost everything.}
		
		\textbf{Tracking}: \textit{Tracking data notifications. Over the years this app has gone down, uninstalled}
		
		\textbf{Privacy Control}: \textit{No privacy. Won't let you ``go offline''. Won’t allow you to disable ad's based on your activity.}
		
	\end{tcolorbox}
	\caption{Review Examples of Different Privacy Concerns}
	\label{fig:privacy_concern_exp}
\end{figure}

\textbf{Answer to RQ\textsubscript{3}}: \textit{Our PCDT algorithm outperforms three topic modeling baselines on detecting privacy concerns from user reviews in terms of topic coherence and topic diversity. Five major privacy concerns are detected from privacy reviews of 8 popular Apps across 4 application categories, including data selling, account security, data security, tracking, and privacy control}.

\section{Threats to Validity and Limitations}


%
%


Threats to internal validity mainly come from the human annotation and evaluation procedure involved in the components of privacy reviews retrieval and classification. To mitigate these, we conduct training sessions with illustrative review examples before starting peer-coding based annotation. The inconsistency annotation instances are resolved through negotiation and majority voting. Moreover, the ground truth used to evaluate different document embeddings is retrieved by \textit{multi-qa-MiniLM}, which also performs best in the final evaluation. This may bring bias to the evaluation result but can not be easily avoided in a starting point of zero labeled data. We note that there may be better document embeddings than \textit{multi-qa-MiniLM}, which could rank more privacy reviews in front of irrelevant ones.

Threats to external validity mainly relate to the diversity of our dataset, which might decrease the generalizability of our approach. To mitigate these, we  select the popular Apps across different common categories based on the App rank lists to ensure that there are sufficient user reviews for evaluating the effectiveness of our data-driven approach~\cite{maalej2015toward}. Furthermore, our reviews are collected from the Apple App Store. The limited number of App categories and single review source might make the findings derived from our research not directly extend to other App categories and review sources, e.g., Google Play.

Threats to construct validity lie in the usage of privacy policies as a query to retrieve candidate privacy reviews. Though privacy policies contain more specific privacy related information compared with general privacy indicating keywords, they mainly state the privacy data practice of App companies. As the main body of privacy policies is about how users' personal data are processed, we thus use the privacy policies as a proxy of user privacy concerns by removing those privacy irrelevant statements.




\section{Related Work}

\subsection{Privacy Requirements Analysis}
RE literature handles privacy requirements mainly from the perspectives of \textit{Compliance}~\cite{breaux2008analyzing, liao2020measuring}, \textit{Access Control}~\cite{caramujo2019rsl}, \textit{Verification}~\cite{fisler2005verification}, and \textit{Usability}~\cite{thomas2014distilling, nguyen2019short, hatamian2019revealing, ebrahimi2021mobile}. \textit{Usability} research focuses on the evaluation and understanding of user behaviors, needs, and motivations through observation techniques, and analysis of usability problems of existing privacy solutions~\cite{anthonysamy2017privacy}. Our work also falls into this perspective. The spectrum of \textit{Usability} perspective is widely ranging from user studies on privacy perceptions~\cite{hatamian2019revealing,tao2020identifying, nema2022analyzing,ebrahimi2022unsupervised} and privacy breaches in social media~\cite{nguyen2019short}, to improvement of user awareness~\cite{wisniewski2017making}.


What distinguish our work from those most related work~\cite{hatamian2019revealing,tao2020identifying, nema2022analyzing,ebrahimi2022unsupervised} is that we propose a novel approach based on text classification and topic modeling to detect common privacy concerns from popular Apps. On the one hand, the user reviews of commonly used Apps have been proven to be an effective strategy for inspiring requirements elicitation~\cite{ferrari2023strategies}. As popular Apps have a huge user base, privacy concerns detected from their reviews could facilitate the privacy requirements elicitation and analysis thus better inform developers Privacy-by-Design~\cite{hustinx2010privacy,bhatia2018empirical}. Furthermore, our approach can be also applied to the review collections of Apps in categories related to the App under consideration so as to detect category-specific privacy concerns, which corresponds to the \textit{similarity by software scope} strategy of App Store-inspired requirements elicitation~\cite{ferrari2023strategies}. On the other hand, our topic modeling results on reviews also confirm the privacy concerns that are identified from large scale reviews on Google Play~\cite{nema2022analyzing}. This verification is necessary for better understanding App users' requirements as there is a difference between the Android and 
iOS versions of hybrid Apps~\cite{malavolta2015end,malavolta2015hybrid} as well as the review culture between different platforms (e.g., Google Play and Apple Store)~\cite{martin2016survey}.

\subsection{App Reviews Mining}

App review mining is another area closely relevant to our study. App stores have long been treated as valuable repositories~\cite{harman2012app} and user reviews posted on them also provide plenty of valuable requirements knowledge~\cite{pagano2013user}. Considerable research efforts have been undertaken to analyze different types of requirements information, e.g., feature requests~\cite{malgaonkar2022prioritizing}, bug reports~\cite{haering2021automatically}, domain-specific topics~\cite{tushev2022domain}, and non-functional qualities~\cite{jha2019mining}. In contrast to those more general themes, we focus on the user privacy concern which is more fine-grained yet significant for user satisfactory~\cite{nguyen2019short}. Our study is complementary to the above existing studies. Applying their methods to the output of our approach could help identify those features, bugs, and quality attributes related to different privacy concerns. Therefore, the combination of our approach and those proposed in previous studies could better inform the maintenance of Apps on privacy aspects.

\section{Conclusion and Future Work}

%
%
%


We propose an App user privacy concerns mining approach composed of candidate privacy reviews extraction, privacy reviews identification, and privacy concern topics detection. Our approach is empirically evaluated on a large collection of App reviews. The evaluation results of a series of experiments demonstrate the effectiveness of our approach.

In the future, we will extend our work to perform much deeper analysis of privacy concerns entailed in the reviews. Several valuable avenues include comparing the privacy concerns among different categories, investigating users' attention on various types of personal data, and detecting potential App privacy issues and/or violations from user reviews.

\newpage

\section*{}
\bibliographystyle{elsarticle-num-names}
\bibliography{pri_topic.bib}

\end{document}